\documentclass[twocolumn,showpacs,preprintnumbers,amsmath,amssymb,floatfix]{revtex4}
\usepackage{graphicx} 
\usepackage{dcolumn}  
\usepackage{textcomp} 
\usepackage{times}

\newcommand{\degree}{\ensuremath{\text{\textdegree}}}
\newcommand{\microm}{\ensuremath{\text{\textmu m}}}
\newcommand\ceil[1]{\left\lceil#1\right\rceil}
\newcommand\floor[1]{\left\lfloor#1\right\rfloor}

\begin{document}


\title{Multi-directional sorting modes in deterministic
lateral displacement devices}

\author{
 Brian R. Long$^{\dag}$\footnote{Current address: Biomedicial Engineering Division, Oregon Health and Science University, 3303 S.W. Bond Ave.
Portland, OR, 97239}, Martin Heller$^{\ddag}$,
 Jason P. Beech$^{\ast}$, Heiner Linke$^{\dag}$,
 Henrik Bruus$^{\ddag}$ and Jonas O. Tegenfeldt$^{\ast}$}


\affiliation{%
$^{\dag}$Materials Science Institute and Department of Physics,
University of Oregon, Eugene, Oregon 97403-1274 USA \\
$^{\ddag}$Department of Micro- and Nanotechnology, Technical
University of Denmark, DTU Nanotech Building 345 East, DK-2800
Kongens Lyngby,
Denmark\\
$^{\ast}$Department of Physics, Division of Solid State
Physics \& Nanometer Consortium, Lund University, S-22100 Lund,
Sweden}

\date{\today}

\begin{abstract}
Deterministic lateral displacement (DLD) devices separate
micrometer-scale particles in solution based on their size using a
laminar microfluidic flow in an array of obstacles.  We investigate
array geometries with rational row-shift fractions in DLD devices by
use of a simple model including both advection and diffusion.  Our
model predicts novel multi-directional sorting modes that could be
experimentally tested in high-throughput DLD devices containing
obstacles that are much smaller than the separation between
obstacles. 
\end{abstract}

\maketitle

\section{Introduction\label{sec:Introduction}}
Deterministic lateral displacement (DLD) is a mechanism of particle
separation that uses the laminar properties of microfluidic flows in
a periodic array of posts to sort particles based on size.  This
technique has been shown to differentiate between
micrometer-sized particles with a resolution in diameter on the
order of 20~nm. The basic sorting mechanism has been described for
the devices used experimentally: particles smaller than a critical
radius $r_\mathrm{c}$ follow streamlines through the array while
larger particles are systematically `bumped' laterally during each interaction with
a post \cite{HuaCoxAus0405,InglisD06,DavIngMor0610}.

Previous analysis of DLD sorting has focused on predicting
$r_\mathrm{c}$ as a function of array parameters, typically the
width of the gap between posts and the shift of posts between rows.
Once basic hydrodynamics is included, theoretical calculations of
$r_\mathrm{c}$ agree with experimental results within about 5~$\%$
\cite{InglisD06, BeeTeg08, helbru08}.  Inclusion of diffusion in DLD
sorting has been described using rough estimations
\cite{HuaCoxAus0405,InglisD06,DavIngMor0610}, and in more detailed
studies that incorporate both microfluidic advection and diffusion
to calculate $r_\mathrm{c}$ under a range of experimental conditions
\cite{helbru08}.

\begin{figure}[!t]
  \centering
  \includegraphics{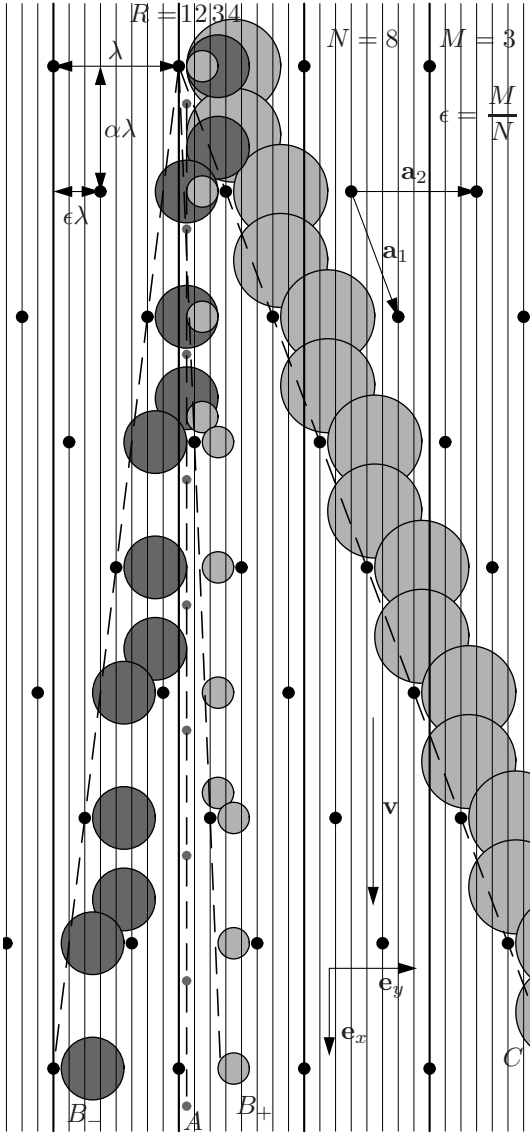}
  \caption{\label{fig:N08M3}%
An array of posts (marked by black dots) with period $N=8$ and a shift of
$M=3$ flow lanes per row, i.e., a row-shift fraction $\epsilon = 3/8
= 0.375$. The flow $\textbf{v}$ is directed along the $x$-axis from
top to bottom. The dashed lines indicate the four possible
separation directions. First, the two well-known modes: the
straight mode $A$ with $\theta_A = 0\degree$ for particles of radius $r$
with $r<(1/N)\lambda$ (small dark gray circles starting in flow lane
$R=1$), and the maximal displacement mode $C$ with $\theta_C =
20\degree$ for $(3/N)\lambda < r$ (large light gray circles starting
in flow lane $R=4$). Additionally, the two novel separation modes:
one $B_+$ towards the right with angle $\theta_{B_+} = 2.4\degree$
for $(1/N)\lambda < r<(2/N)\lambda$ (small light gray circles
starting in flow lane $R=2$), and another $B_-$ towards the left
with angle $\theta_{B_-} = -7.1\degree$ for $(2/N)\lambda <
r<(3/N)\lambda$ (large dark gray circles starting in flow lane
$R=3$). The solid vertical lines indicate the flow lanes of width $\lambda/N$, while $\mathbf{a}_1$ and $\mathbf{a}_2$ are the lattice vectors,
 and $\alpha$ is an aspect ratio.}
\end{figure}

Previous analysis of the geometry of the DLD array has been limited
to the following conventional case. In a given row the
center-to-center distance between the posts is denoted $\lambda$, see Fig.~\ref{fig:N08M3}.
The subsequent row of posts is placed at a distance $\alpha\lambda$
downstream from the first row. Normally, $\alpha$ is chosen to be
unity, however this is not an essential requirement. The posts in
this second row are displaced a distance $(1/N)\:\lambda$ along the
row, where $N$ traditionally has been an integer. The ratio $1/N$ is also denoted the
row-shift fraction $\epsilon$. In row number $N+1$ the posts have
the same positions as in the first row, and consequently the array
is cyclic with period $N$. Due to this periodicity of the array and
the laminarity of the flow, the stream can naturally be divided into
$N$ flow lanes, each carrying the same amount of fluid flux, and each
having a specific path through the device \cite{HuaCoxAus0405}.

For devices with the simple row-shift fraction $\epsilon = 1/N$ and
disregarding particle diffusion, only one critical separation size
$r_\mathrm{c}$ is introduced. Spherical particles with a radius
smaller than $r_\mathrm{c}$ will move forward along the main
flow direction through the device, defining the angle $\theta = 0$.  However, particles with a radius larger than $r_\mathrm{c}$ are forced by collisions
with the posts to move in a skew direction at an angle
$\theta$ given by $\tan\theta = 1/(\alpha N)$. Taking diffusion into
account the transition from straight to skew motion takes place over a
finite range of particle sizes \cite{HuaCoxAus0405,InglisD06,DavIngMor0610,helbru08}.

In this paper we generalize the array geometry by studying the
effects of row-shift fractions different from that of the
conventional, simple $(1/N)$-array. We show in
Sec.~\ref{sec:BasicTheory} that by displacing consecutive rows by
the rational fraction $\epsilon \lambda =  (M/N)\lambda $, where $M$ is an integer that
is not a divisor of $N$, two new separation modes appear, each
associated with a distinctive range of particle sizes and separation
directions $\theta$. Furthermore, to test experimental
feasibility of the novel separation modes, we introduce in
Sec.~\ref{sec:Model} a model of the DLD system reduced to its
essential elements: particle trajectories interrupted by
size-dependent interactions with a periodic array of posts.
Utilizing these simplifications, we investigate in
Sec.~\ref{sec:Results} the advection and diffusion of particles in
the $M/N$-array geometries, and discuss in Sec.~\ref{sec:Discussion}
possible experimental consequences of our novel DLD system.

In our model of the DLD system described in Sec.~\ref{sec:Model} we
reduce the posts to point-like obstacles in a uniform flow. This
particular case is currently of interest to researchers looking to
apply DLD separation to high-throughput microfluidic devices. Such a
reduced post size decreases hydraulic resistance and thus increases
the liquid throughput for a given pressure difference applied along
the device. One promising method to create such devices is to use
arrays of semiconductor nanowires \cite{MarBorSei0312} in a
microfluidic channel.

\section{Basic theoretical analysis}
\label{sec:BasicTheory}
The introduction of a non-simple row-shift fraction $\epsilon = M/N$
in the DLD system is first discussed in
Sec.~\ref{sec:SpecificTheory} for the specific case of $M/N = 3/8$,
since all the novel separation modes are present in that device geometry.
Fig.~\ref{fig:N08M3} shows the principle of the fractionally
displaced DLD array leading to multi-directional separation of
particles of different sizes. As was the case for the simple row
shift fraction $1/N$, the rational row-shift fraction $M/N$ also
naturally leads to $N$ flow lanes, each carrying the same amount of
fluid flux.  In this section, all particles are assumed to follow these flow lanes unless bumped by an interaction with a post.  However, in contrast to the traditional DLD geometries, now the posts are displaced $M$ flow lanes instead of just a single flow lane when passing from one row of
posts to the next. 

In Sec.~\ref{sec:GeneralTheory} we analyze this more general
case of $M/N$-arrays, where the integer row-shift $M$ and the integer
array-period $N$ have no common divisors.

\subsection{The specific row-shift fraction 3/8}
\label{sec:SpecificTheory}
First we consider the explicit choice of parameters given in
Fig.~\ref{fig:N08M3}, namely, a period $N=8$, and a row-shift of
$M=3$ lanes in the $y$-direction $\mathbf{e}_y$, i.e., a row-shift
fraction of $\epsilon = 3/8$. The flow $\mathbf{v}$ is in the
$x$-direction $\mathbf{e}_x$. For simplicity, we employ the most
simple model where all flow lanes are assumed to have the same width
$\lambda/N$, and where the particles are not subject to Brownian
motion. The analysis can straightforwardly be extended to take the
different widths of the flow lanes~\cite{InglisD06} as well as
diffusion~\cite{helbru08} into account.

The analysis is most easily carried out by considering spherical
particles of increasing radius $r$. As the rows in
Fig.~\ref{fig:N08M3} are shifted to the right, it is natural to
choose the starting point of a given particle to be directly to the right of a
post, placing the particle's center in flow lane $R = 1, 2, 3$, or $4$ according to size.

For the smallest particles with $r<\lambda/N$, labeled $A$ in Fig.~\ref{fig:N08M3}, we obtain a path
corresponding to the familiar so-called zigzag path defined in
Ref.~\cite{HuaCoxAus0405}. Due to the point-like nature of our obstacles, the path is a straight line, indicated by the dashed vertical
line in Fig.~\ref{fig:N08M3}. The path angle is $\theta_A = \arctan
0 = 0\degree$.

For the next set of particles with $\lambda/N<r<2\lambda/N$,
($B_+$ in Fig.~\ref{fig:N08M3}),
we note that they are not affected significantly by passing the
second rows of posts. The displacement of $(M/N)\lambda$ is larger
than the size of the particle. By simple inspection we find that the
particles interact with a post in the fourth row leading to a bump
of one lane width to the right. This bumping brings the particles
back to a position just right of a post, and we have identified a
new separation mode, $B_+$.  The direction of mode $B_+$ can
be characterized by the integers
\begin{subequations}
  \begin{align}
    p &= \parbox[t]{7cm}{\raggedright the number of rows after which the bumping pattern repeats itself, and} \\
    q &= \parbox[t]{7cm}{\raggedright the number of flow lanes that the particles are bumped  to the right while traveling through $p$ rows.}
  \end{align}
\end{subequations}
Here, with $p = 3$ and $q = 1$ and the array parameters indicated in
Fig.~\ref{fig:N08M3}, the path angle of mode $B_+$ is found to be
$\theta_{B_+} = \arctan\big[1/(\alpha\times3\times8)\big] =
2.4\degree$. Here and in the following we choose the aspect ratio
$\alpha = 1$.

For the third set of particles with $2\lambda/N<r<3\lambda/N$,
marked as $B_-$ in Fig.~\ref{fig:N08M3},
we note that they collide with a post in the second row and
are bumped two lanes to the left. After two rows, the particles are again bumped two lanes to the left,
and we have identified another new separation mode, $B_-$.
Given this period $p=2$ bumping of $q=-2$ flow lanes (where minus
indicates displacement to the left), the path angle of mode $B_-$ is
found to be $\theta_{B_-} = \arctan\big[
-2/(\alpha\times2\times8)\big] = -7.1\degree$.

\begin{table}[t]
  \centering
  \newcolumntype{R}{D{r}{\;r}{6,6}}
  \newcolumntype{Q}{D{/}{/}{-1}}
  \newcolumntype{S}{D{.}{.}{-1}}
  \newcommand\Bm{B_{\rlap{\scriptsize$-$}}}
  \newcommand\Bp{B_{\rlap{\scriptsize$+$}}}
  \begin{ruledtabular}
    \caption{\label{tab:sepparameters} List of separation radii $r$ and angles
    $\theta$ as a function of the integer array parameters $N$, $M$, $p$ and $q$
    for $\alpha=1$.}
  \noindent
  \begin{tabular*}{0.9\linewidth}{@{\extracolsep{\fill}}*{3}{>{$}c<{$}}RQS}
    N & M & \multicolumn{1}{c}{mode}
          & \multicolumn{1}{p{3cm}}{\centering
              particle radius in\\units of lane width}
   &  q/p & \multicolumn{1}{p{2.5cm}}{\centering
              separation angle\\
              $\theta=\arctan\big[\tfrac{p}{qN}\big]$\rule[-5pt]{0pt}{5pt}}\\\hline
   5 & 2 & A   & 0<r<1   &  0/1 &  0.0\degree \\
     &   & \Bm & 1<r<2   & -1/2 & -5.7\degree \\
     &   & C   & 2<r<2.5 & 2/1  & 21.8\degree \\ \hline
   7 & 2 & A   & 0<r<1   & 0/1  &  0.0\degree \\
     &   & \Bm & 1<r<2   & -1/3 & -2.7\degree \\
     &   & C   & 2<r<3.5 & 2/1  & 15.9\degree \\ [2mm]
   7 & 3 & A   & 0<r<1   & 0/1  &  0.0\degree \\
     &   & \Bm & 1<r<3   & -1/2 & -4.1\degree \\
     &   & C   & 3<r<3.5 & 3/1  & 23.2\degree \\ \hline
   8 & 3 & A   & 0<r<1   & 0/1  &  0.0\degree \\
     &   & \Bp & 1<r<2   & 1/3  &  2.4\degree \\
     &   & \Bm & 2<r<3   & -2/2 & -7.1\degree \\
     &   & C   & 3<r<4.0 & 3/1  & 20.6\degree \\ \hline
   9 & 2 & A   & 0<r<1   & 0/1  &  0.0\degree \\
     &   & \Bm & 1<r<2   & -1/4 & -1.6\degree \\
     &   & C   & 2<r<4.5 & 2/1  & 12.5\degree \\ [2mm]
   9 & 4 & A   & 0<r<1   & 0/1  &  0.0\degree \\
     &   & \Bm & 1<r<4   & -1/2 & -3.2\degree \\
     &   & C   & 4<r<4.5 & 4/1  & 24.0\degree \\ \hline
  10 & 3 & A   & 0<r<1   & 0/1  &  0.0\degree \\
     &   & \Bm & 1<r<3   & -1/3 & -1.9\degree \\
     &   & C   & 3<r<5.0 & 3/1  & 16.7\degree \\ \hline
  11 & 2 & A   & 0<r<1   & 0/1  &  0.0\degree \\
     &   & \Bm & 1<r<2   & -1/5 & -1.0\degree \\
     &   & C   & 2<r<5.5 & 2/1  & 10.3\degree \\ [2mm]
  11 & 3 & A   & 0<r<1   & 0/1  &  0.0\degree \\
     &   & \Bp & 1<r<2   & 1/4  &  1.3\degree \\
     &   & \Bm & 2<r<3   & -2/3 & -3.5\degree \\
     &   & C   & 3<r<5.5 & 3/1  & 15.3\degree \\ [2mm]
  11 & 4 & A   & 0<r<1   & 0/1  &  0.0\degree \\
     &   & \Bp & 1<r<3   & 1/3  &  1.7\degree \\
     &   & \Bm & 3<r<4   & -3/2 & -7.8\degree \\
     &   & C   & 4<r<5.5 & 4/1  & 20.0\degree \\ [2mm]
  11 & 5 & A   & 0<r<1   & 0/1  &  0.0\degree \\
     &   & \Bm & 1<r<5   & -1/2 & -2.6\degree \\
     &   & C   & 5<r<5.5 & 5/1  & 24.4\degree \\ \hline
  12 & 5 & A   & 0<r<1   & 0/1  &  0.0\degree \\
     &   & \Bp & 1<r<2   & 1/5  &  1.0\degree \\
     &   & \Bm & 2<r<5   & -2/2 & -4.8\degree \\
     &   & C   & 5<r<6.0 & 5/1  & 22.6\degree
  \end{tabular*}
  \end{ruledtabular}
\end{table}

Finally, the fourth set of particles (with $3\lambda/N<r$) is
considered, shown as the large light gray circle in
Fig.~\ref{fig:N08M3}. Since $3\lambda/N$ equals the row-shift
$\epsilon\lambda$, these large particles collide with a post in each
row ($p=1$) where they are bumped $q = M = 3$ lanes to the right.
This is the conventional maximal displacement mode $C$ \cite{HuaCoxAus0405}. As a result
the path angle for mode $C$ here is found to be $\theta_C =
\arctan\big[3/(\alpha\times8)\big] = 20.6\degree$.

\subsection{\boldmath General row-shift fractions $M/N$}
\label{sec:GeneralTheory}
In the general case of a DLD device with period $N$ and a row-shift
of $M$ flow lanes, it is useful to introduce the floor function
$\lfloor x \rfloor$ of $x$, which gives the largest integer smaller
than or equal to $x$, eg., $\lfloor8/3\rfloor = 2$ and
$\lfloor10/3\rfloor = 3$, and the ceiling function $\ceil{x}$ of $x$
which gives the smallest integer larger than or equal to $x$ (see
also the definitions given at Ref.~\cite{floorfct}).

Using the notation in Fig.~\ref{fig:N08M3},  the flow lane $R$
occupied by the center of the particles can be expressed in terms of
the particle radius $r$ as $R = \ceil{rN/\lambda}$, so that $R=1, 2,
3, \ldots, \ceil{N/2}$ for~$0<r<\lambda/2$.

Two cases are straightforward to analyze. For small radii with
$R=1$, the particles will follow the streamlines without any
systematic net lateral displacement, i.e., a mode $A$ in the
direction $\textbf{t}_A$ given by
 \begin{equation} \label{eq:partAdir}
 \textbf{t}_A = \alpha\: \textbf{e}_x,
 \end{equation}
and forming the path angle $\theta_A$ with the $x$-axis,
 \begin{equation}
 \theta_A = 0,
 \quad R=1.
 \end{equation}
For large radii with $M<R<\ceil{\frac{N}{2}}$, the particles collide
with the posts and are bumped $M$ flow lanes to the right in each
row, but they do not get stuck between the posts; this is mode $C$.
The path is directed along the direction $\textbf{t}_C$ given by
 \begin{equation} \label{eq:partCdir}
 \textbf{t}_C = \alpha\: \textbf{e}_x + \frac{M}{N}\:\textbf{e}_y,
 \end{equation}
and forming the path angle $\theta_C$ with the $x$-axis,
 \begin{equation}
 \theta_C = \arctan\bigg[\frac{M}{\alpha N}\bigg],
 \quad M<R<\ceil{\frac{N}{2}}.
 \end{equation}
In a given $M/N$ array, modes with larger sorting angles are
excluded because of the post spacing in the $y$-direction: particles
with radius $r>\lambda/2$ are unable to fit between the posts.

If the particles are small enough to pass the second row without getting bumped to the right, but too large for mode $A$, $1<R\leq M$, their trajectories fall into one or two $B$ modes.

\begin{figure}[phtb]
  \centering
  \includegraphics{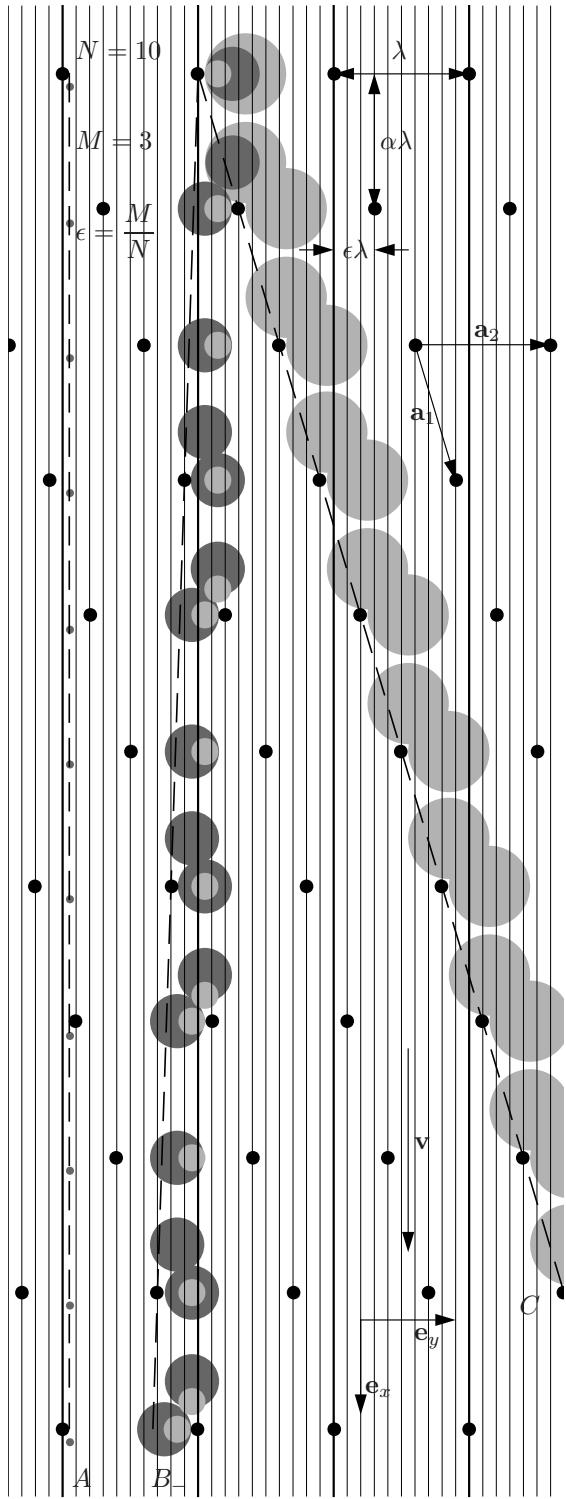}
  \caption{\label{fig:geometry}
  An array with period $N=10$ and a shift of $M=3$ flow lanes per row, giving a row-shift fraction $\epsilon = 3/10 = 0.3$. The flow $\textbf{v}$ is directed along the $x$-axis from top to bottom. Here there are three sorting modes, delimited by two critical radii. Mode $A$ for particles of radius $r$ with $r< r_{\mathrm{c}1} = (1/10)\lambda$ (shown on the far left), and mode $C$, the maximal displacement mode  for $(3/10)\lambda = r_{\mathrm{c}2}< r< (1/2) \lambda$ (large light gray circles).  A novel mode $B_-$ displaces particles with $r_{\mathrm{c}1}<r<r_{\mathrm{c}2}$ towards the left (large dark gray and intermediate, light gray circles).}
\end{figure}

As a particle is convected through the array, a post will approach
the particle from the left in steps of $M$ flow-lanes per row the
particle advances, hence the use of modulus $M$ arithmetic in the
following analysis.

If $(N \bmod M)<R\leq M$ the particle will hit the post with its
center to the left of this obstacle and will therefore enter mode
$B_-$ where it is displaced to the left with a period
$p_-=\floor{\frac{N}{M}}$. This is most readily seen by starting the
analysis with a particle position just left of a post. A particle
with $(N\bmod M)<R\leq \bigl[(N-R+1)\bmod M\bigr]$ will bump left
after $p_-=\floor{\frac{N}{M}}$ rows and will again be in a position
just left of a post. The small particle in mode $B_-$ of
Fig.~\ref{fig:geometry} is an example of this behavior.  Slightly
larger particles with $\bigl[(N-R+1)\bmod M\bigr]<R\leq M$ will bump
right after $p=\floor{\frac{N-R+1}{M}}$ rows. Since we are only
considering particles with $R\leq M$, this displacement will always
be less than $M$ flow lanes, and the particle is therefore bound to
bump left on the post in the following row, i.e., after a total of
$p_-=\floor{\frac{N}{M}}$ rows. The large $B_-$ mode particle in
Fig.~\ref{fig:geometry} is an example of this behavior.

The trajectories in mode $B_-$ have period $p_- =
\floor{\frac{N}{M}}$. The number $q_-$ of lanes bumped after passing
these $p_-$ rows is $q_- = Mp_- - N < 0$. The path is directed along the direction $\mathbf{t}_{B_{-}}$ given by
 \begin{subequations}
 \begin{align} \label{eq:partBMdir}
 \textbf{t}_{B_-} &= \alpha p_-\:\textbf{e}_x +
 \frac{q_-}{N}\textbf{e}_y,\\
 p_- &= \floor{\frac{N}{M}},\\
 q_- &= Mp_- - N < 0,
 \end{align}
 \end{subequations}
forming the path angle $\theta_{B_-}$ with the $x$-axis,
 \begin{equation}
 \theta_{B_-} = \arctan\bigg[\frac{q}{\alpha p_- N}\bigg] < 0.
 \end{equation}

If $1<R\leq(N \bmod M)$ the particle will enter mode $B_+$ where it
is displaced to the right with a period $p_+=\floor{\frac{N+R}{M}}$.
To realize this it is natural to start the analysis with the
particle just right of a post. Again, a post will approach the
particle from the left in steps of $M$ lanes as the particle moves
through the array. A particle with $1<R\leq\ceil{\frac{M}{2}}$ will
follow the flow for $p_+=\floor{\frac{N+R}{M}}$ rows and then bump
right. If $\ceil{\frac{M}{2}}<R\leq(N\bmod M)$ the particle will
bump left already in the second row of posts. The particle is now in
a position just left of a post. However, since it is not large
enough to follow the $B_-$ path, it will bump right when it meets
the post after $p_+=\floor{\frac{N+R}{M}}$ rows.

The trajectories in mode $B_+$ have period $p_+ =
\floor{\frac{N+R}{M}}$. After $p_{+}$ rows the particles will get bumped $q_{+}$ flow lanes to the right given by $q_+ = Mp_+ - N>0$.  The path
is directed along the direction $\mathbf{t}_{B_{+}}$ given by
 \begin{subequations}
 \begin{align} \label{eq:partBPdir}
 \textbf{t}_{B_+} &= \alpha p_+\:\textbf{e}_x +
 \frac{q_+}{N}\textbf{e}_y,\\
 p_+ &= \floor{\frac{N+R}{M}},\\
 q_+ &= Mp_+ - N > 0,
 \end{align}
 \end{subequations}
forming the path angle $\theta_{B_+}$ with the $x$-axis,
 \begin{equation}
 \theta_{B_+} = \arctan\bigg[\frac{q_+}{\alpha p_+ N}\bigg] > 0.
 \end{equation}

In terms of the flow lane number $R$, the criteria for the four
different displacement modes can be summarized as follows
\begin{subequations}
\begin{align}
  & \text{mode $A$,}   & \quad & \text{if $R = 1$}                       \\
  & \text{mode $B_+$,} &       & \text{if $1<R\leq(N\bmod M)$}             \\
  & \text{mode $B_-$,} &       & \text{if $(N\bmod M)< R\leq M$}             \\
  & \text{mode $C$,}   &       & \text{if $M<R\leq\ceil{\tfrac{N}{2}}$.}
\end{align}
\end{subequations}
Note that mode $B_+$ vanishes if $(N\bmod M)=1$.

\section{Model and Implementation\label{sec:Model}}
The following model is established to numerically test the sorting behavior
of a particular $M/N$ DLD array and take into account the effect of particle diffusion on sorting behavior, as discussed below.  We treat the device as a periodic array of
zero-radius posts with the geometry shown in
Fig.~\ref{fig:geometry}.  This $N=10$, $M=3$ geometry, with a row-shift
fraction given by $\epsilon = 3/10$, exhibits the three modes shown in Table~\ref{tab:sepparameters}, including a novel sorting mode, $B_-$.


We assume the array to be infinitely deep so that the flow field is
two-dimensional and independent of the $z$-direction. Consistent with
the infinitesimal size of the posts, the liquid flow through the
device is assumed to be uniform with velocity $\textbf{v} =
v\:\textbf{e}_x$ along the $x$-axis. Thus, our model does not
describe Taylor-Aris dispersion, which in real systems with
finite-sized posts would be induced along the $x$-direction by a
combination of transverse diffusion and transverse velocity
gradients \cite{Bruus2008}. The particles only interact with the
posts through a hard-wall repulsion and any effect of the particles on fluid flow
is neglected. The particle-post interaction excludes the center of a
particle with radius $r$ from a circular region of the same radius
around the point-sized post.  In addition to being moved by the
fluid and interacting with the posts, each particle has a diffusion
coefficient $D$ given by the Einstein relation
\begin{equation} \label{eq:Ddef}
D(r) = \frac{k_\mathrm{B}T}{6\pi\eta r},
\end{equation}
where $k_\mathrm{B}T$ is the thermal energy and $\eta$ is the
viscosity of the water. For the calculations below we have chosen
the following experimentally relevant parameters: For water at room
temperature $k_\mathrm{B}T \approx 4 \times 10^{-21}$~J and $\eta
\approx 10^{-3}$~Pa$\:$s, and for the geometry the post separation
is $\lambda = 10~\microm$ and particle radii in the range
$0.5~\microm < r < 4~\microm$.

A final basic assumption of our model is that all time dependence in
our model is implicitly given by the advective flow speed $v$.
For particles starting at the entrance of the device at $x=0$ the
time $t$ is given through its $x$-coordinate as $t=x/v$. The model
therefore allows all the relevant dynamics of an ensemble of many
particles to be described by a continuous concentration distribution
$c(x,y)$ with some given initial distribution $c(0,y)$ at the
entrance of the DLD device. Given $c(0,y)$ the time-evolution of the
distribution consists of calculating $c(\Delta x,y)$ after
convection to $x = \Delta x$. By following the evolution of $c(x,y)$
as the distribution interacts with posts and responds to thermal
forces, our model can identify the basic modes of transport in an
array of posts and the effect of diffusion on this transport. 

\begin{figure}[t]
  \centering
  \includegraphics{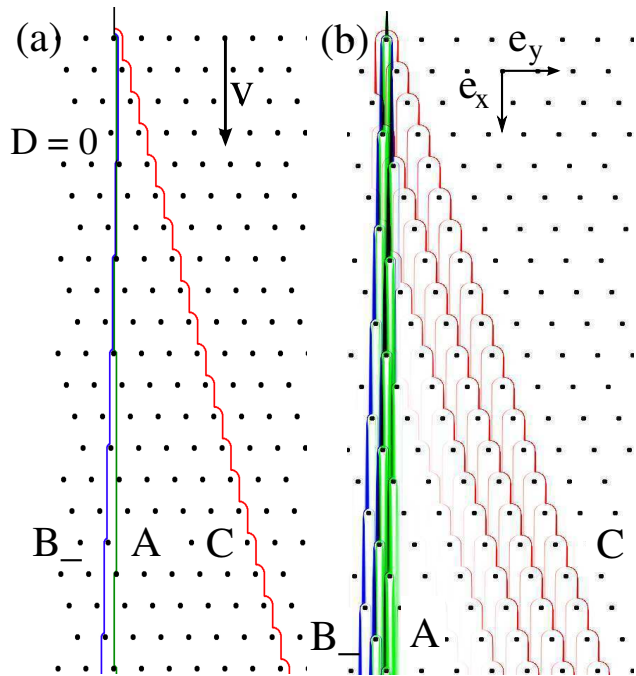}
  \caption{\label{fig:Modes}%
 [Color online]  Composite image of numerically calculated spatial distributions for a device with $N=10$ and $M=3$. (a) Results for $D=0$ for particles with radii $r < r_{\mathrm{c}1}$ (mode $A$), $r_{\mathrm{c}1}< r \leq r_{\mathrm{c}2}$ (mode $B$) and $r  > r_{\mathrm{c}2}$ (mode $C$). (b) Particles, with the same radii as in (a), moving through the array with a flow speed of $v = 100$~\microm/s, including the effect of diffusion.  Broadening of all distributions due to diffusion can be seen and particles in mode $C$ are sorted less efficiently.  Initial spatial distributions here are the same for all particle radii, and a narrow initial distribution is used for visual clarity.}
\end{figure}

The initial distribution $c(0,y)$ is given by a box
distribution of width $\lambda$ (although a narrow distribution is used in Fig.~\ref{fig:Modes} for visual clarity), and the distribution $c(\Delta x,y)$ is
calculated from the previous distribution $c(0,y)$ taking into
account its interactions with the posts as well as the diffusion
equation.  The entire distribution $c(x,y)$ is evaluated by iterating the following procedure:
\begin{enumerate}
\item Upon encountering a row of posts, the distribution for particles of radius $r$ is set to zero in regions with a distance smaller than $r$ to any post, and the corresponding number of particles is then added to the distribution in the adjacent pixels to maintain the total number of particles (see Fig.~\ref{fig:Modes}).
\item The distribution $c(x,y)$ is subsequently evolved in accordance with the diffusion equation, with the diffusion coefficient given by Eq.~(\ref{eq:Ddef}),
    \begin{equation} \label{eq:diffusion}
      v \frac{\partial c}{\partial x}
      = D \frac{\partial^{2} c}{\partial y^{2}},
    \end{equation}
    employing the implicit time $t = x/v$ set by convection along the $x$-direction, and using the Fourier cosine transformation in the transverse $y$-direction as described below.
\end{enumerate}

The computation uses a finite array of width $w = 10\lambda$, i.e.\ containing 10 posts, and the row separation is again taken to be equal to the post separation, i.e.\ $\alpha =1$.
The array with width $w$ is discretized in $y$ into
$n_\mathrm{max}=10^4$ pixels of size $\Delta w \times \Delta x$ with
$\Delta x = \Delta w = w/n_\mathrm{max}$.

The discrete Fourier cosine transformation $C(x,k_n)$ of the
distribution $c(x,y)$ then takes the form
 \begin{equation} \label{eq:Cdef}
 C(x,k_n) = \frac{2-\delta_{0,n}}{w}\:
 \int_0^w c(x,\tilde{y})\:\cos(k_n \tilde{y})\:\mathrm{d}\tilde{y},
 \end{equation}
where $k_n$ is given by
 \begin{equation} \label{eq:kndef}
 k_n = \frac{2\pi}{w}\:n,\quad n=0,1,2,\ldots,n_\mathrm{max} =
 \frac{w}{\Delta w}.
 \end{equation}
By direct inspection we find the well-known result from the
$y$-dependent diffusion equation in Fourier space that, during the
time step $\Delta t =\Delta x/v$, $C(x,k_n)$ evolves into
$C(x+\Delta x,k_n)$ as
 \begin{equation} \label{eq:Cdiffusion}
 C(x+\Delta x,k_n) = C(x,k_n)\:
 \exp\bigg[-Dk_n^2\frac{\Delta x}{v}\bigg].
 \end{equation}
By the inverse Fourier cosine transform we can therefore write the
distribution at row $x+\Delta x$ in terms of that at row $x$ as
 \begin{equation} \label{eq:clambda}
 c(x+\Delta x,y) = \sum_{n=0}^{n_\mathrm{max}}
 C(x,k_n) \:\exp\bigg[-Dk_n^2\frac{\Delta x}{v} \bigg]
 \:\cos(k_n y).
 \end{equation}
which by construction automatically respects the boundary condition that no particles can diffuse beyond the edges of the array.  The evolution of the distribution due to diffusion is computed at each row of pixels after the effects of posts on the distribution have been taken into account.

To elucidate the sorting mechanism in the absence of thermal forces,
calculations were also done with diffusion coefficient $D=0$, in
which case $c(x,y)$ evolves only according to the interaction of the
particles with the posts.  Results of these calculations are shown
as $D=0$ in Fig.~\ref{fig:avr}.

While $c(x,y)$ is the calculated distribution at a given time and
position in the array, the set of all $c(x,y)$ also represents the
steady-state distribution of a stream of particles entering an array
of obstacles and moving constantly through the array, as seen in
Fig.~\ref{fig:Modes}.

The calculations were done using Matlab on a personal computer and a
64-bit dual processor workstation.

\subsection{Results\label{sec:Results}}

\subsubsection{Three transport modes in the 3/10-array}
The existence of the novel sorting mode $B_-$, as well as the two modes $A$ and $C$ previously
described in DLD literature are confirmed by applying our numerical model to a range of particle sizes advected through the 3/10-array. As the particle distributions move
through the array, their trajectories form three modes $A$, $B$ and
$C$, according to two critical radii, $r_{\mathrm{c}1}$ and
$r_{\mathrm{c}2}$, see Fig.~\ref{fig:Modes}(a). Our calculations
reproduce the two known modes: the `zigzag mode' $A$, in which there
is no average displacement from the direction of flow, and the
`bumped mode' $C$, in which particles are bumped laterally in every
row. These two modes are most clearly seen in
Fig.~\ref{fig:Modes}(a), where the distributions are calculated
without diffusion. In mode $A$, where $r \leq r_{\mathrm{c}1} =
(1/N)\lambda$, particles may interact with the posts, but no net
lateral displacement is accomplished.  Mode $C$ is characterized by
a displacement equal to the shift $(M/N)\lambda$ for every row the
particles pass through. In the novel mode $B_-$, particles of size $
r_{\mathrm{c}1} < r \leq r_{\mathrm{c}2}$ interact with posts more
frequently than in mode $A$ but less frequently than in Mode $C$, as
described in Secs.~\ref{sec:GeneralTheory}.  The 3/10 array used
here clearly exhibits the lone $B_-$ mode shown in
Table~\ref{tab:sepparameters} for these array parameters.  It is
important to note that mode $B_-$ vanishes in the conventional case $M=1$, and all particles smaller than the critical radius $r_{\mathrm{c}2}$ move along the direction of flow.

The directions $\mathbf{t}_A$ and $\mathbf{t}_C$ of the conventional modes $A$ and $C$ are given directly by Eqs.~(\ref{eq:partAdir}) and~(\ref{eq:partCdir}) for $\alpha = 1$:
 \begin{subequations}
 \begin{align}
 \textbf{t}_A &= \textbf{e}_x,\\
 \textbf{t}_C &= \textbf{e}_x + \frac{3}{10}\:\textbf{e}_y,
 \end{align}
while the direction $\mathbf{t}_{B_{-}}$ of mode $B_-$ is found through the path period
$p_-  = \floor{\frac{N+R}{M}} = \floor{\frac{10+1}{3}} = 3$ and the
lane shift $q_- = p_-M - N = 3\times3 - 10 = -1$, and thus
 \begin{equation}
 \textbf{t}_{B_-} = 3\:\textbf{e}_x - \frac{1}{10}\:\textbf{e}_y.
 \end{equation}
 \end{subequations}
The corresponding displacement angles become
 \begin{subequations}
 \begin{alignat}{4} \label{eq:displacementangles}
 \theta_A   &=& 0.0\degree,& & \quad
   0\:\frac{\lambda}{10} <&\;r\;&<1\: \frac{\lambda}{10},\\
 \theta_B{_-} &=&\; -1.9\degree,& & \quad
   1\: \frac{\lambda}{10} <&\;r\;&<3\: \frac{\lambda}{10},\\
 \theta_C   &=& 16.7\degree,& & \quad
   3\: \frac{\lambda}{10} <&\;r\;&<5\: \frac{\lambda}{10}.
 \end{alignat}
 \end{subequations}

The array parameters used here can be translated into those used in DLD literature \cite{HuaCoxAus0405,InglisD06,DavIngMor0610}, simply by setting $M=1$.

\subsubsection{Effect of diffusion on sorting}
\begin{figure}[t]
  \centering
  \includegraphics{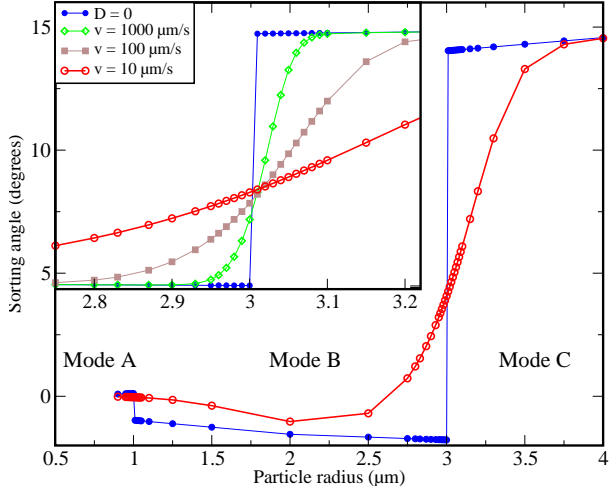}
  \caption{\label{fig:avr}%
 Sorting angles $\theta$ calculated for a device with $N=10$ and $M = 3$ as described in Sec.~\ref{sec:Model} with $r_{\mathrm{c}1}$ = 1~\microm\ and $r_{\mathrm{c}2}$ = 3~\microm.  The initial position corresponds to the center of mass of the initial distribution between two posts and the final position corresponds to the center of mass after 10 rows. $\theta$ is plotted here versus particle radius $r$ with (open red circles) and without (filled blue circles) diffusion with flow velocity $v=10$~\microm/s.  The negative sorting angles for $r_{\mathrm{c}1} < r < r_{\mathrm{c}2}$ indicate the presence of mode $B_-$ for this array.  Inset shows the sorting angle $\theta$ around $r = r_{\mathrm{c}2}= 3$~\microm\ for a range of flow velocities (same $y$-axis range). Diffusion blurs the sharp transition between the sorting modes, as discussed in Sec.~\ref{subsec:diffdiscussion}.}
\end{figure}

The effect of diffusion on the sorting of particles is shown in
Fig.~\ref{fig:avr}.  The angles shown are measured between
$\mathbf{v}$ and the lateral displacement of the center of mass of the distribution for each
particle size after 10 rows of posts for high and low flow speeds.
We can estimate speeds at which diffusion becomes negligible by
comparing the time  it takes a particle to be advected along the
$x$-direction from one row to the next, $\lambda/v$, to the time it
takes a particle to diffuse transversely in the $y$-direction to
reach a position where it would be bumped,
$2D/(r-r_{\mathrm{c}2})^2$. For high flow speeds,
 \begin{equation}
 v \gg \frac{2D\lambda}{(r-r_{\mathrm{c}2})^{2}},
 \end{equation}
diffusion can be neglected, and the transitions between the sorting
modes are sharp, as seen in the $D=0$ case.  Note that this
velocity diverges as the particle size approaches the critical radius
$r_{\mathrm{c}2}$; in this limit the displacement needed for a
particle to change sorting directions goes to zero.  Within the
spatial resolution of this work (1~pixel = 10~nm), the particles
closest in size to the critical radius will still be sensitive to
diffusion at flow velocities below 10~mm/s.  As flow speeds
decrease, particles have more time to diffuse transversely as they
move through the array, and the effects of thermal motion on sorting
are seen more clearly. Transverse diffusion of particles along the
$y$-direction tends to shift the center of mass of the distribution
$c(x,y)$ towards the midpoint between posts. This means that
particles with $r < r_{\mathrm{c}_2} = \frac{ 3 \lambda}{10}$ are more likely to be shifted
to higher sorting angles. However, in the regions between rows,
diffusion allows particles to move transversely away from the path
that would normally be `bumped' by a post, decreasing their sorting
angle. These two effects of diffusion are responsible for the
smoothing of the angle versus radius curves for slower flow speeds
in Fig.~\ref{fig:avr}.  The calculated values for $\theta_B{_-}$ are in good agreement with the value predicted in Eq.~(\ref{eq:displacementangles}), but for particles with $r > r_{\mathrm{c}_2} = \frac{ 3 \lambda}{10}$, the finite width of the initial distribution and the relatively short array size (10 rows) reduce the calculated values for $\theta_C$ from the predicted value by about $15~\%$.  The small variation in sorting angle with radius for modes $B_-$ and $C$ for $D=0$ in Fig.~\ref{fig:avr} is mainly the result of the two end points used to define the angle being not exactly equivalent: the position of the second, but not the first end point varies continuously with bead size, and so the presented angle varies with bead size.  Secondly, since the number of rows is not divisible by the periodicity of mode $B_-$, an additional small error is introduced.  These deviations should vanish for simulations with larger numbers of rows.


\section{Discussion} \label{sec:Discussion}

\subsection{The novel sorting mode and its relation to kinetically locked-in transport}
\label{sec:discussmode} DLD devices have thus far been made with a
fixed flow direction and almost exclusively with $M=1$. However, in
theoretical work studying transport through periodic potential
landscapes, the direction of the applied force is varied for a fixed
array geometry and the transport direction is calculated
\cite{ReiNor9901, A.M0609,LadKasGri0407, RoiWonGri0701}.  To calculate the
correspondence between varying the array parameters $M$ and $N$ used
here and changing the flow direction in a fixed array as in \cite{ReiNor9901, A.M0609,LadKasGri0407, RoiWonGri0701} is cumbersome,
but for a range of flow directions near $\textbf{t}_{B_{-}}=3\textbf{e}_x -
\frac{1}{10}\:\textbf{e}_y$, the angles to the flow direction
$\theta_{B_-}$ and $\theta_{C}$ vary as the flow directions change,
but the relative angle between them, $\theta_{C}-\theta_{B_-}$,
remains a constant defined by the array. The angle between modes
$B_-$ and $C$ is insensitive to small changes in flow
direction for $\mathbf{v}$ near (in this case) $3\textbf{e}_x -
\frac{1}{10}\:\textbf{e}_y$.

This insensitivity to flow direction is an example of a plateau in a
so-called `devil's staircase': transport through a 2-D periodic
potential is independent of the flow direction near small integer lattice vectors
\cite{ReiNor9901}. In this case the lattice vectors are
$\textbf{a}_1 = \textbf{t}_{C}=\textbf{e}_x + (3/10)\:\textbf{e}_y$ and
$\textbf{a}_2 = \textbf{e}_y$, and the two close-lying flow
directions are $\textbf{t}_{B_{-}}= 3\textbf{e}_x - \frac{1}{10}\:\textbf{e}_y =
3\:\textbf{a}_1 - \textbf{a}_2$ and $\textbf{a}_1 =\textbf{t}_{C}$.

The interplay between lattice directions and applied forces has been
documented extensively in the literature of kinetically and
statistically locked-in transport.  Of interest in the present
context is that many numerical simulations of trajectories through
various two-dimensional periodic potentials have been done to study
these and other phenomena, including sorting of particles
\cite{ReiNor9901, A.M0609,LadKasGri0407, RoiWonGri0701}.

The interaction between posts and particles that we have chosen
simplifies DLD to a 1D distribution that evolves in time.  This
allows the effects of diffusion to be easily incorporated into our
modeling of the dynamics of the distribution of particles.  Also,
the particular interaction between point-sized posts and
finite-sized particles depends only on particle size, an analysis
that seems to be absent from the literature.

\subsection{Diffusion, detectability and experimental possibilities \label{subsec:diffdiscussion}}
A clear difference between the results in Fig.~\ref{fig:avr}, based
on zero-sized posts, and those reported in the literature, based on
finite-sized posts, is that the critical radius (defined as the inflection point of the angle \textit{vs.} radius graph near $r = r_{c}$), decreases
for lower flow velocities in Ref.~\cite{HuaCoxAus0405}, whereas
our simulations show a critical radius that is essentially constant.
When particles have more time to diffuse laterally in reported
experimental data, ones that previously followed the `zigzag' path
follow something closer to Mode $C$ but not the other way around. We
have identified the difference in size of the posts as the primary
basis for the difference in symmetry. In the gap between the posts,
only beads smaller than $r_\mathrm{c}$ can change modes (from $A$ to
$C$) whereas beads larger than $r_\mathrm{c}$ cannot change modes
because of steric hindrance. Diffusion between posts is thus
asymmetric. On the other hand, between rows all beads can change
modes equally well so that the effect of diffusion is symmetric.
This result is most clearly seen in two cases: (\textit{i}) with
sufficiently large posts and small spacing between rows, diffusion
between posts dominates leading to asymmetry, and (\textit{ii}) with
our needle-like posts, instead diffusion between the rows dominates,
leading to symmetry between small and large particles. In devices
with large round posts such as those in
Ref.~\cite{HuaCoxAus0405}, the flow streams are narrower in the
gap between the posts than in the region between the rows making the
asymmetric diffusion even more pronounced. The symmetry about
$r_{\mathrm{c}2}$ shows that sorting in this model is robust against
changes in flow velocity.

As discussed in Sec.~\ref{sec:discussmode}, there is no difference
between modes $A$ and $B_-$ when the flow is directed along the
lattice direction $3\textbf{a}_1-\textbf{a}_2$, which is equivalent
to a conventional array with $M=1$ and $N=3$, instead of along
$\mathbf{e}_x$. Also, while mode $B_-$ for the 3/10 array shown in
Fig.~\ref{fig:geometry} is directed away from mode C, the mode $B_+$ discussed in
Sec.~\ref{sec:SpecificTheory} is deflected away from $\mathbf{v}$
towards mode $C$. The absence of modes $B_-$ and $B_+$ in previous
analyses of DLD experiments stems from the use of tilted square
arrays with flows chosen such that $M=1$ or more general arrays that
are still limited to simple row-shifts $1/N$. In these cases, Modes
$A$ and $B$ are the same: they both go along the direction of flow.
Interestingly, in their paper Ref.~\cite{InglisD06}, Inglis
\textit{et al.} mention that they are studying simple row-shift
fractions $\epsilon = 1/N$, with $N$ being an integer, but they do
not comment on the data points in their Fig.~2 that clearly have $\epsilon \neq
1/N$.

\begin{figure}[t]
  \centering
  \includegraphics{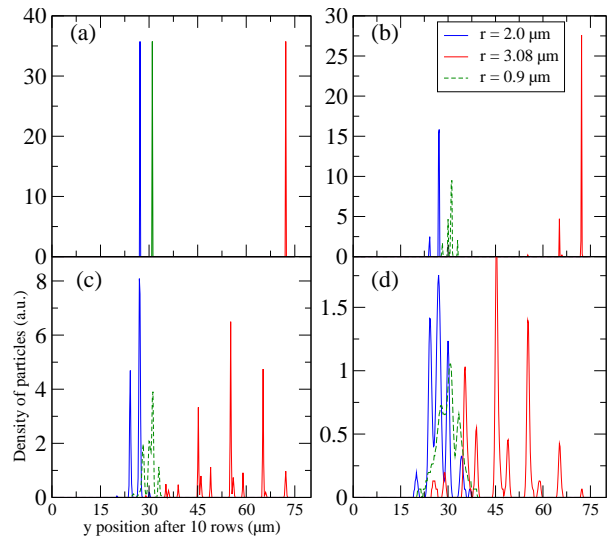}
  \caption{\label{fig:spatialwidths} [Color online] Distributions of three particle sizes: $r = 0.90$ (green dashed line), $2.00$ (blue), and $3.08$~\microm\ (red) after transport through ten rows of the $N = 10$, $M=3$ array.  The total number of particles is the same in each case and each initial distribution (not shown) is a square distribution with a narrow width centered on $y = 30~\microm$.  (a) No diffusion. (b) With diffusion and $v = 1000$~\microm/s. (c) With diffusion and $v = 100$~\microm/s. (d) With diffusion and $v = 10$~\microm/s. Panel (b) shows a case where mode $B_-$ could be detected experimentally.  For the lower speed in panel (c), modes $A$ and $B_-$ cannot be resolved, but the combined distribution is broader than mode $A$ alone.  For the even lower speed in panel (d) the distributions of particles in modes $A$ and $B_-$ are each wider than the separation between them and the two modes are completely unresolvable.}
\end{figure}

Experimental detection of mode $B$ requires that the distributions
of modes $A$ and $B$ must be spatially separated.  The numerically calculated distributions shown in
Fig.~\ref{fig:spatialwidths} exhibit four qualitative regimes that
could be observed in an experiment to detect the presence of
particle transport in mode $B$.
\begin{enumerate}
\renewcommand\theenumi{(\alph{enumi})}
\item
  At very high flow speeds, corresponding to $D = 0$ in the numerical data, the three modes are completely separated because each distribution is very narrow.  In this regime, arbitrary spatial separation can be achieved simply by running the particles through a longer
  array.
\item
  At high intermediate flow speeds, the distributions have widened due to diffusion, but modes $A$ and $B$ are clearly distinguishable, despite some overlap.
\item
  At low intermediate flow speeds, modes $A$ and $B$ overlap enough to prevent resolution of two separate distributions.  This regime is relevant to DLD device design because it would be experimentally observed as an anomalous, asymmetric broadening of the distribution associated with the `zigzag' path.
\item
  At low flow speeds, distributions from modes $A$ and $B$ are completely overlapping and it may even be difficult to differentiate them from mode $C$.
\end{enumerate}

Experimental realization of the regime investigated in this model would require arrays made with very small posts to minimize hydrodynamic effects on particle trajectories.  This also corresponds to a reduction in hydrodynamic drag, which is beneficial for researchers seeking to increase fluid throughput of devices.

As can be seen in Fig.~\ref{fig:Modes}, the angle $\theta_{B}$ is small compared to $\theta_{C}$.  In order to differentiate between particles traveling in modes $A$ and $B$, size dispersion of beads must be considered in addition to broadening due to diffusion. Commercially available polystyrene beads used in DLD experiments typically have size distributions with widths of less than $\pm$10~\%.  This then requires choosing particles whose size distributions are separated by more than 10~\%, such as those shown in Fig.~\ref{fig:spatialwidths}, or the use of a DLD array to create a sufficiently narrow size distribution.  If hydrodynamic effects or limitations on flow velocity in a particular experiment prevent the novel sorting mode from being completely resolved, it may still appear as an asymmetric broadening of the distribution of seemingly undeflected particles, as in Fig.~\ref{fig:spatialwidths}.

In general, the separation angles for a given M/N-array can be made larger to the extent that the aspect ratio $\alpha$ can be made smaller without risking clogging of the largest particles. By consulting Table~\ref{tab:sepparameters}, it can be seen that the novel separation angle of the 3/10-array is one of the smaller $B$ angles, and also, the 3/8-, 3/11-, 4/11- and 5/12-arrays offer both the $B_+$ and the $B_-$ modes.

\section{Conclusions\label{sec:Conclusions}}
We have identified novel sorting modes in a model of transport
through a DLD device characterized by row-shift fractions $M/N$. Our
simple model also reproduces key features of DLD arrays, including
sorting based on size and the blurring of cutoffs between modes due
to diffusion. Even if not completely resolved, the novel sorting
mode has the potential to increase spatial broadening of `zigzag'
particle distributions.  In order to avoid this broadening,
adjustable DLD arrays could use variable spacing while maintaining a
fixed $M =1$ geometry, such as in Ref.~\cite{BeeTeg08}, or tune flow
angles to exactly reproduce the $M = 1$ condition across a fixed
obstacle array using techniques such as in Ref.~\cite{jotref}. Our
simulations indicate that using needle-like posts decreases the
shift in critical size due to diffusion that has been observed in devices where the
post separation is on the same scale as the post diameter.
Furthermore, the use of more general array geometries and simplified
fluid dynamics links this work to the field of kinetically locked
transport phenomena.

\section{Acknowledgements}
This research is supported by the National Science Foundation under CAREER Grant No.~0239764 and IGERT International Travel Award, the Danish Research Council for Technology and Production Sciences Grant No.~26-04-0074, and the Swedish Research Council, under Grants No.~2002-5972 and 2007-584.

\end{document}